\documentclass[12pt]{iopart}
\newcommand{\ds}{\displaystyle}

\begin{document}
\title[Logarithmic perturbation theory for radial Klein-Gordon equation]
{Logarithmic perturbation theory for radial Klein-Gordon equation
 with screened Coulomb potentials
via $\hbar$ expansions}
\author{I V Dobrovolska and R S Tutik}
\address{Department of Physics,
Dniepropetrovsk National University, Dniepropetrovsk, 49050,
Ukraine}

\ead{tutik@ff.dsu.dp.ua}

\begin{abstract}
The explicit semiclassical treatment of logarithmic perturbation
theory for the bound-state problem within the framework of the
radial Klein-Gordon equation with attractive real-analytic screened
 Coulomb potentials, contained time-component of a
Lorentz four-vector and a Lorentz-scalar term, is developed.
Based upon $\hbar$-expansions and suitable quantization
conditions a new procedure for deriving perturbation expansions
is offered. Avoiding disadvantages of the standard approach, new
handy recursion formulae with the same simple form both for
ground and excited states have been obtained. As an example, the
perturbation expansions for the energy eigenvalues for the
Hulth\'en potential containing the vector part as well as the
scalar component are considered.
\end{abstract}

\submitto{\JPA} 

\maketitle

\section{Introduction}Static screened Coulomb potentials have been
widely used in nuclear and particle physics, atomic physics,
solid-state physics and chemical physics. Usually the bound-state
problem with these potentials is considered within a
non-relativistic framework. Nevertheless, relativistic effects
for a particle under the action of such potential could become
important, especially for strong coupling.

 Several attempts have been made to describe
relativistic systems in a central field due to a time-component of
 Lorentz four-vector
and a Lorentz-scalar interaction within the framework of the
Klein-Gordon equation.
However, for almost all potentials this
equation is not exactly solvable which compels to resort to some
approximation methods.

A number of such approaches to solving the radial Klein-Gordon
equation in analytical expressions have been developed,
including, in particular, the use of the relativistic hypervirial
and Hellmann-Feynman theorems \cite{b1},  the $1/N$-expansions
[2-9], the method of Regge trajectories [10-12], the elements of
an $SO(2,1)$ Lie algebra \cite{b11, b12} and a perturbation
scheme based on a comparison equation \cite{b13}.

Despite such a variety of methods one of the most popular
techniques is still logarithmic perturbation theory [16-18].
Within the framework of this theory, the conventional way to
solve the quantum-mechanical bound-state problem consists in
changing from the wave function to its logarithmic derivative and
converting the Klein-Gordon equation into the nonlinear Riccati
equation. In the case of ground states, the consequent expansion
in a small parameter leads to a hierarchy of simple equations
that permit us to derive easily the corrections to the energy and
wave function for each order. However, when radially excited
states are considered, the standard approach \cite{b16} becomes
extremely cumbersome and, practically, inapplicable for
describing higher orders of expansions. At the same time the
evaluation of perturbative terms of large orders is needed for
applying modern summation procedures because the obtained series
are typically divergent.

The above mentioned weakness of the standard approach is caused
by factoring out zeros of the wave functions with taking into
account corrections to the positions of the nodes. On the other
hand, the radial quantum number, $n$, that is equal to the number
of nodes of the wave function, most conveniently and naturally is
introduced in the consideration by means of quantization
conditions as well as in the WKB-approach \cite{b17, b18}.
However, since the WKB-approximation is more suitable for
obtaining energy eigenvalues in the limiting case of large
quantum numbers but perturbation theory, on the contrary, deals
with low-lying levels, the WKB quantization conditions need
change.

Recently, a new technique based on a specific quantization
conditions has been proposed to get the perturbation series via
the semiclassical $\hbar$-expansions within the one dimensional
Schr\"{o}dinger equation \cite{b19,b91}.

In this paper, we would like to extend similar technique to the
bound-state problem for the Klein-Gordon equation with the
Coulomb-type potential, contained time component of a Lorentz
 four-vector and/or a Lorentz scalar term, that receives attention
as possible model of quark confinement.

There are several advantages to described approach: no wave
functions or matrix elements need to be previously calculated and
derived recursion formulae for obtaining perturbation corrections
have the same simple form both for ground and excited states.

The proposed technique can be regarded as a further investigation
of a part assigned to a rule of achieving a classical limit for
 radial and orbital quantum numbers in
the construction of the semiclassical methods for the
Klein-Gordon equation. If the WKB-approach is realized using the
rule, $\hbar\rightarrow0$, $n\rightarrow\infty$,
$l\rightarrow\infty$, $\hbar n = const$, $\hbar l = const$,
application of the condition $\hbar\rightarrow0$, $n = const$,
$l\rightarrow\infty$, $\hbar n \rightarrow 0$, $\hbar l = const$
was proved to lead to the method of $1/N$-expansion \cite{b7,b71}
and the method of Regge trajectories \cite{b9, b10}. Here we
address ourselves to the alternative possibility:
$\hbar\rightarrow 0$, $n = const$, $l = const$, $\hbar n
\rightarrow 0$, $\hbar l \rightarrow 0$, that results in the
explicit semiclassical treatment of the logarithmic perturbation
theory for the radial Klein-Gordon equation.

This paper is organized as follows: section II contains a general
discussion of the method and necessary assumptions in
semiclassical treatment of the logarithmic perturbation theory.
In section III quantization conditions obtained are used for
deriving the recursion relations for perturbation expansions.
Section IV demonstrates that described approach restores the exact
results in the case of the Coulomb potential and gives the
example of its explicit application to the bound-state problem in
field of the Hulth\'en potential. This potential, apart from
its initial interest as a possible form of nuclear interaction,
 is exactly solvable for the $l=0$ states, thereby providing a
 consistency check for our perturbation technique.
 Finally we present a brief
summary.

\section{The method}
In this section, we study the bound state problem for a relativistic
scalar
particle moving in the field of an attractive central
real-analytic screened Coulomb potential, admitted bounded
 eigenfunctions, having in consequence a discrete energy spectrum.
  This potential has a Coulomb-like behaviour at the origin, caused
  by the time component of a Lorentz four-vector, $V(r)$, and/or a
Lorentz-scalar term, $W(r)$, which in general can be written as
\begin{equation} \label{2.1}
V(r) =\frac{1}{r}{ \sum_{i=0}^{\infty}{V_i}\,r^i}\;\;\;\;\;\;
W(r)=\frac{1}{r}{ \sum_{i=0}^{\infty}W_i\,r^i\; }.
 \end{equation}
Notice, that in the last case, when the Coulombic singularity is
contained only in one part of potential, another part may be a
smooth function described by equation (\ref{2.1}) with $ V_0=0$
or $ W_0=0 $.

In what follows, a scalar potential will be included in the mass
term $ m(r) $, by analogy with "dynamical mass" models of quark
confinement \cite{b20,b21}
\begin{equation}\label{2.2}
 m(r)=m+ \frac{ W(r) }{ c^2 }  \; .
 \end{equation}
Then the reduced radial part of the Klein-Gordon equation
takes the form
\begin{equation}\label{2.3}
\hbar^2R''(r)=\left\{m^2(r)c^2-\frac{1}{c^2}[E-V(r)]^2+\frac{\hbar^2
l(l+1)}{r^2}\right\}R(r)\;
\end{equation}
where the particle wave function is $\Psi (r)=R(r)/r$.

As is customary in the logarithmic perturbation theory, we apply
the substitution, $ C(r)= \hbar R'(r) / R(r) $, and go over from
(\ref{2.3}) to the Riccati equation
\begin{equation}\label{2.4}
\hbar C'(r)+C^2(r) =\frac{\hbar^2 l(l+1)}{r^2}
+m^2(r)c^2-\frac{1}{c^2}[E-V(r)]^2\;.
\end{equation}

Within the framework of the standard approach to the logarithmic
perturbation theory, all quantities of the Riccati equation are
expanded in formal series in powers of a small screening parameter
$\lambda$, incoming in a potential functions of the form
$V(r)=r^{-1}f(\lambda\,r)$. After such an expansion the screening
parameter appears in consideration in combination with
derivatives of screening function as $\lambda^i\, \mathrm{d}
^if/\mathrm{d} r^i$.Taking into account this fact, in our approach
we do not single out the screening parameter, but incorporate it
into coupling constants $V_i$ and $W_i$ from (\ref{2.1}).

Besides, because the Riccati equation (\ref{2.4})
has in the relativistic case the same structure as in the
nonrelativistic one
\cite{b19,b91}, these coupling constants, $V_i$ and $W_i$,
 appear in common with powers of Plank's constant.
Therefore the perturbation series must be in reality not only
expansions in powers of a screening parameter but also the semiclassical
$\hbar$-expansions, too. As stated earlier, we intend to restore the results of
logarithmic perturbation theory for the $n$th eigenfunction and
corresponding energy eigenvalue by means of the explicit
expansions in powers of $\hbar$.

For this purpose we apply the method of series expansions in
small parameter at the higher derivative, known from the theory
of differential equations \cite{b22}. Such an approach does not
imply knowledge of the exact solution for initial approximation
which is obtained automatically. Only the leading order in $\hbar$
 for the energy eigenvalues need be determined first. With
 assumption that the energy eigenvalues and the logarithmic
 derivative of the wave function may be written as an asymptotic
 power series in the Planck constant, from (\ref{2.4}) we then have
\begin{equation}\label{2.5}
E= \frac{1}{\hbar^2}  \sum_{k=0}^{\infty}E_k\hbar^{2k}\;
\end{equation}
\begin{equation}\label{2.6}
C(r)= \frac{1}{\hbar}  \sum_{k=0}^{\infty}C_k(r) \hbar^{2k}\, .
\end{equation}
Here we take into account that under substitution into
equation(\ref{2.4}) the coefficients of odd powers of $\hbar$ for
the energy expansion (\ref{2.5}) and even powers for (\ref{2.6})
are equal to zero.

Before proceeding, a few words about the order in $\hbar$ of the
combination $\hbar c$, appeared, as readily seen from
(\ref{2.4}), in our consideration. For velocities which are small
in relation to the velocity of the light $c$, relativistic
mechanics must go over into classical mechanics. So the
non-relativistic energy $E_{NR}$ should be regarded as small
quantity in comparison to the relativistic rest energy $mc^2$.

Because for the Coulomb interaction, which is treated as the dominant
part of the screened Coulomb potentials, the non-relativistic
quantum-mechanical energy behaves as $E_{NR}\sim1/\hbar^2$, from the
inequality $E_{NR}<<mc^2$ we conclude that $\hbar c\sim O(1)$. It means
that the factor $\hbar c$ is merely the coefficient providing the
right dimension.

Therefore, to avoid additional complications, in foregoing
consideration we put
$\hbar c=1 $. It is worth noting that analogous statement for
transition to classical limit is hold within the framework of the
Dirac equation, too \cite{b23}.

Further, through the use of the $\hbar$-expansions (\ref{2.5}) and
(\ref{2.6}), on collecting coefficients of the same powers of
$\hbar$, from the Riccati equation (\ref{2.4}) we obtain

\begin{eqnarray}\label{2.7}
C_0^2(r)  =  m^2-E_0^2\; \nonumber \\
C_0(r)C_1(r)  =  E_0\Bigl[  V(r)-E_1 \Bigr]  + m W(r) \; \nonumber \\
C_1'(r)+2C_0(r)C_2(r)+ C_1^2(r)   = \frac {l(l+1)}{r^2}
-E_1^2-2E_0E_2+2E_1V(r) \\
 \quad +W^2(r)-V^2(r)\; \nonumber \\
 \cdots  \nonumber \\
 C_{k-1}'(r)+\sum_{j=0}^kC_j(r)C_{k-j}(r)
  =  -\sum_{j=0}^kE_jE_{k-j}+2E_{k-1}V(r) \;\; k>2\;.\nonumber
\end{eqnarray}

In the case of ground states, the recurrence system at hand
coincides with one derived by means of the standard technique and
can be solved straightforwardly. It will be recalled that
complications of the logarithmic perturbation theory arise in the
description of radial excitations when the nodes of wavefunctions
are separated in some factor. We intend to circumvent this
difficulty by making use of the quantization condition. Its
fundamental idea that stems from the
 WKB-approach \cite{b17,b18} is well known as the principle of argument in
the analysis of complex variables. Being applied to the logarithmic
 derivative of wave function it means that
\begin{equation}\label{2.8}
\frac{1}{2\pi\,\rm{i}}\oint{C(r)\,{\rm d}
r}=\hbar\,n\;\;\;\;\;n=0,1,2,...
\end{equation}
where $n$ is a number of nodes of the wave function,
$R(r)$, and a contour of integration encloses only these nodes.

This condition is exact and is widely used for deriving the
high-order corrections to the WKB-approximation \cite{b18} and the
1/$N$-expansions \cite{b7,b71}. Now one important detail must be
noted. The radial and orbital quantum numbers, $n$ and $l$,
correspondingly, are specific quantum
 notions and need be defined before going over from quantum
mechanics to classical physics. Therefore the quantization
condition (\ref{2.8})
 must be supplemented with the rule of achieving a classical limit that
stipulates the type of semiclassical approximation.

It should be stressed that in our approach the quantization
condition is quite distinguished from the WKB condition (\ref{2.8})
in two points: we apply the different rule of achieving a classical
limit and choose another contour of integration in complex plain.

Firstly, about the rule of achieving a classical limit.

Unfortunately, under the influence of standard textbooks on quantum
mechanics, the semiclassical $\hbar$-expansions are usually associated
solely with the WKB-approach, for which the rule of achieving a
classical limit is
\begin{equation}\label{2.9}
  \hbar\to 0,\;n\to\infty,\; l\to\infty,
\;\hbar\,n={\rm const},\;\hbar\,l={\rm const}.
\end{equation}
At that time the semiclassical 1/$N$-expansion, being complementary to the
 WKB-method, requires the conditions \cite{b7,b71}
\begin{equation}\label{2.10}
  \hbar\to 0,\; n={\rm const},\;l\to\infty,
\;\hbar\, n\to 0,\;\hbar\,l={\rm const}.
\end{equation}

As it will be shown, the semiclassical treatment of logarithmic
perturbative theory proved to involve the alternative possibility:
\begin{equation}\label{2.11}
  \hbar\to 0,\; n={\rm const},\;l={\rm const},
\;\hbar\,n\to 0,\;\hbar \,l\to 0.
\end{equation}

We should remark that the part of this rule concerned the orbital
 quantum number,
 $l$, differs from one used within the WKB-approach and the
  1/$N$-expansion
method. In our consideration this part, implying the first order
in $\hbar$ for the quantity $\hbar l$, has been used in deriving
the system (\ref{2.7}).

The second distinction is concerned with the choice of the path of
integration in complex plane.In contrast to the WKB-approach and
the 1/$N$-method, we choose such a contour of integration which
encloses not only the nodes of the wave function but the boundary
point, $r=0$, as well, and no other singularities.

As it follows from equation (\ref{2.3}), the regular branch of the
function $R(r)$ behaves in a neighborhood of the origin as
\begin{equation}\label{2.12}
R(r)_{r\to 0}\sim r^{1/2+\sqrt{{W_0^2-V_0^2+(l+1/2)^2}}}U(r)\; ,
\end{equation}
where $U(r)$ contains only nodes of the wave function. Hence, the
direct integration of the logarithmic derivative $R'(r)/R(r)$ around
the origin yields the quantity $1/2+\sqrt{{W_0^2-V_0^2+(l+1/2)^2}}$
 in addition to nodes in the quantization condition (\ref{2.8})
 which can be now rewritten as
\begin{equation}\label{2.13}
  \frac{1}{2\pi\,\rm{i}}\oint{C(r)\,{\rm d} r}=\hbar\left({n+1/2+\sqrt{{W_0^2-V_0^2+(l+1/2)^2}}}\right)
\;\;\;n=0,\,1,\,2...
\end{equation}

Taking into account that due to the rule (\ref{2.11}) the
rihgt-hand side of this equality has the first order in $\hbar$,
and on substituting the expansion (\ref{2.6}), our quantization
condition takes its final form
\begin{equation}\label{2.14}
  \frac{1}{2\pi\,\rm{i}}\oint{C_i(r)\,{\rm d} r}=\left(n+1/2+\sqrt{{W_0^2-V_0^2+(l+1/2)^2}} \right)\,\delta_{i,1} \;
\;\;n=0,\,1,\,2...
\end{equation}
where the Kronecker delta $\delta_{ij}$ is used and the quantity,
$W_0^2-V_0^2+(l+1/2)^2$, under square root
sign should be not negative from the constraint to obtain a regular
bound state wavefunction.

 A further application of the theorem of residues to the
explicit form of functions $C_i(r)$ easily solves the problem of
the description of both ground and radially exited states.

\section{Recursion formulae for perturbation expansions}
We proceed now to deriving the recursion relations for obtaining
the $n$th eigenfunction and corresponding energy eigenvalue.

Let us consider the system (\ref{2.7}) and investigate the
behaviour of the functions $C_k(r)$. From the first equation we
have
\begin{equation}\label{3.1}
 C _0(r)=-\sqrt{m^2-E_0^2}\;,
\end{equation}
where the minus sign is chosen from boundary condition. Then the
function $C_1(r)$ has a simple pole at the origin, owing to the
Coulombic behaviour incoming in one or/and other part of the
potential at this point, while the function $C_k(r)$ has a pole
of the order $k$. Hence $C_k(r)$ can be represented by the
Laurent series
\begin{equation}\label{3.2}
  C_k(r)=r^{-k}\sum_{i=0}^\infty{C^{k}_{i}r^i}\;,\; k\geq 1\; .
\end{equation}
With definition of residues, this expansion permits us to express
the quantization conditions (\ref{2.12}) explicitly in terms of
the coefficients $C_i^k$ as
\begin{equation}\label{3.3}
  C_k^{k+1}=N \,\delta_{0,k} \;
\;\;n=0,\,1,\,2...\; \;
\end{equation}
where $N=n+1/2+\sqrt{{W_0^2-V_0^2+(l+1/2)^2}}$.

Thereby the common consideration of the ground and excited states
has been indeed proved to be possible.

The substitution of the series (\ref{3.2}) into the system
(\ref{2.7})
 and collecting coefficients of the like powers of $r$ leads
 to the recursion relation in terms of the Laurent coefficients,
 $C_i^k$:
\begin{eqnarray}\label{3.4}
  C^k_i  & = -\frac{1}{2C_0^0} \biggl[
     (i-k+1)C^{k-1}_i+ \,
     \sum^{k-1}_{j=1}  \sum^i_{p=0} C^j_p C^{k-j}_{i-p}\nonumber
\\
    & + \,\delta_{k,i} \sum^k_{j=0} E_j  E_{k-j} -2 E_{k-1}
    V_{i-k+1}\nonumber
\\
      & + \,\delta_{k,2}  \sum^i_{p=0} (V_p  V_{i-p}- W_p  W_{i-p})
       -\delta_{i,0} \delta_{k,2}\,l ( l + 1 )
\\
& - \,\delta_{k,1} \, 2 m W_i \biggr]\; \nonumber
\end{eqnarray}
where for universality of designations we put $C_0^0=C_0(r)$,
$C^0_i=0$, $i>0$.

In the case $i\not=k$, this formula is intended for obtaining
$C_i^k$, whereas if $i=k$, by equating the explicit expression
for $C^{k+1}_k$ to the quantization condition (\ref{3.3}) we
arrive at the recursion formulae for the energy eigenvalues
\begin{eqnarray}\label{3.5}
  E_0  & = \frac{ m }{N^2+V_0^2}
\left(  N   \sqrt{N^2 + V_0^2 - W_0^2 } - V_0  W_0   \right)
 \;\nonumber
\\
 E_k  & = \frac{C^0_0 C^1_0}{2(E_0 C^1_0+V_0 C^0_0 )} \Biggl\{
      \frac{1} {C^1_0}
      \Biggl[
             \sum^{k-1}_{j=2}\sum^k_{p=0} C^j_p
             C^{k+1-j}_{k-p}\nonumber
  \\
 & + \,2   \Theta   ( k-2 )  \sum_{p=1}^k C^1_p C^{k}_{k-p}
 +\delta_{k,1}  (V_0 V_1- W_0 W_1)
   \Biggr]
\\
 & -\, \frac{1}{C_0^0}
     \Biggl[
       C^{k-1}_k
       +\sum^{k-1}_{j=1}\sum^k_{p=0}C^j_p  C^{k-j}_{k-p}
        +\sum^{k-1}_{j=1} E_j E_{k-j}\nonumber
  \\
& - 2E_{k-1} V_{1} +\,\delta_{k,2}\sum^2_{j=0}(V_j  V_{2-j}-
W_jW_{2-j})\nonumber
     \\
     &  -\delta_{k,1} 2 m W_1
\Biggr] \Biggr\}\, \;k\geq0\,\nonumber
\end{eqnarray}
where we use the step function
\[
\begin{array}{ll}
\Theta(k)&=1,\quad k\geq0\,\\
              &=0,\quad k<0\,
\end{array}
\]
and the plus-sign for $E_0$ is chosen since the particle states
enter to the bound state gap from the upper continuum.

Thus, the problem of finding the energy-eigenvalues and
eigenfunctions for the bound-state problem within the framework
of the radial Klein-Gordon equation with screened Coulomb
potentials can be considered solved. The equations (\ref{3.4}) and
(\ref{3.5}) have the same simple form both for ground and exited
states and define a useful procedure of successive calculation
of higher perturbation theory orders convenient both for analytic
evaluation and numerical computation by a computer.

\section{Discussion and examples}
In this section we consider specific application of our technique
to perturbed hydrogenic problems. But first of all, in the proof
of correctness of the recurrent formulae obtained, we check
whether the exact results for the pure Coulomb potentials are
restored.
 From equations (\ref{3.5}) it is readily seen that
$E_0$ does be exact solution to the Klein-Gordon equation with
time-component of Lorentz four-vector and  Lorentz-scalar
Coulomb potentials \cite{b24,b1}. As an
additional check, we demonstrate that in this case our technique
restores the exact solution for the eigenfunctions, too.

Changing the variable from $r$ to $\rho$, with
$\rho=2\,r\,(m^2-E^2)^{1/2}=2\,\mu\,r$, and putting for
simplicity, $\gamma=\sqrt{(l+1/2)^2+W_0^2-V_0^2},\; \hbar=c=1$,
from the system (\ref{3.4}) we find that $C_k(\rho)=d_k\rho^{-k}$,
where
\begin{eqnarray}\label{4.1}
 d_0 & = -\frac{1}{2}\;\nonumber \\[7pt]
 d_1 & = n+\frac{1}{2}+\gamma \;\nonumber \\[7pt]
 d_2& = n\,(n+2\gamma) \; \\[7pt]
& \cdots \nonumber \\
 d_k & = (1-k)\,d_{k-1}+\sum_{j=1}^{k-1}d_jd_{k-j} \,\;\;\; k>2\; .\nonumber
\end{eqnarray}

Integration of the functions $C_0(\rho)$ and $C_1(\rho)$ gives
exactly the part of the unnormalized radial wave function
\begin{equation}\label{4.2}
R_l(\rho)=e^{-\rho/2}\,\rho^{1/2+\,\gamma}P_n(\rho)\;
\end{equation}
which provides the regular behaviour at the origin and correct one
at infinity.

The remaining part is a polynomial,
 $P_n(\rho)= \sum_{k=0}^{n}a_k \rho^{k}$, that satisfies the
 equation
\begin{equation}\label{4.3}
P'_n(\rho)/P_n(\rho)=\sum_{k=1}^{\infty}d_k \rho^{-k}-
    \left(\frac{1}{2}+\gamma\right)\rho^{-1}\;
\end{equation}
and whose coefficients are determined by the system
\begin{equation}\label{4.4}
a_k (n-k) +\sum_{j=k+1}^{\infty}a_j d_{j-k+1}=0\; .
\end{equation}
The combination of this equations, multiplied by a suitable $d_j$
with a view to taking into account equations (\ref{4.1}), arrives
at the following relation between two consecutive coefficients
\begin{equation}\label{4.5}
\frac{a_{m-1}}{a_m}=\frac{(m+1)(m+2\gamma)}{m-n-1}\;
\end{equation}
that is the recursion formula for the associated Laguerre
polynomials $L^{2\gamma}_n(\rho)$ (see, for example, \cite{b25}).
Thus, described technique restores the exact result for the wave
functions \cite{b24}, too.

Now, as an example of specific application of our technique we
consider energy eigenvalues for the attractive Hulth\'en
potential, having not only time-component of the Lorentz-vector,
 $V(r)=-\frac{a\lambda}{e^{\lambda r}-1}$, but the Lorentz-scalar
 term,
$W(r)=-\frac{b\lambda}{e^{\lambda r}-1}$, as well. Because this
screened Coulomb potential admits of the exact solution only in
the case of $s$-wave \cite{b26,b27}, for describing orbital
excitations we can use the perturbation theory. Then the
coefficients of expansions (\ref{2.1}) determining potentials are
\begin{equation}\label{pot}
\begin{array}{rl}
V_0  & = -a \\[7pt]
V_k  & = \ds -\sum_{j=0}^{k-1}{\frac{V_j\lambda^{k-j}}{(k+1-j)!}}
\,\,\,\;\; k > 0\;\\[7pt]
W_0 & = -b \\[7pt]
W_k & = \ds -\sum_{j=0}^{k-1}{\frac{W_j\lambda^{k-j}}{(k+1-j)!}}
\,\,\,\;\; k > 0\;
\end{array}
\end{equation}
and with applying equation (\ref{3.5}) the analytic expressions
for perturbation corrections to the bound state energy take the
form

\begin{equation}\label{4.6}
\begin{array}{rl}
  E_0  & \ds  = \frac{ m}{N^2+a^2}
\left(  N   \sqrt{N^2 + a^2 - b^2 } - a \, b   \right)
 \;
\\[7pt]
E_1 & \ds   ={\frac{\lambda \, }{2\,m}\left( {a}\,m +
       {b}\,{{E }_0} \right)}\;
\\[7pt]
E_2 & \ds   = - {\frac{{{\lambda }^2}\,( {b}\,m + {a}\,{{E }_0}
)}{24\,
     ( {a}\,m + {b}\,{{E}_0} ) \,
     \mu^2 \,m}}
\,
     \left[3\,(a\,m + b E_0)^2 - \mu^2 \, L \right]\;
\\[7pt]
E_3 & \ds   =-\frac{\lambda\, b}{2\,m} E_2\;
\\[7pt]
E_4 & \ds   =- \frac{ {\lambda }^4\,\left( {b\,m + a\, E_0}
\right)}
    {5760\,{\left( a\,m + b\, E_0 \right)}^3\,{\mu }^4\,m^3}\,
\\ & \ds \times
    \left\{
     45\left[ {\left( a\,m + b\, E_0 \right)}^6
     +5\,b^2\mu^2{\left( a\,m + b\, E_0 \right)}^4 \right]\right.\\[7pt]
     & \ds  \left.
     +L\,\left[ 24\,a^3m^3b\,E_0\left(6\,E_0^2-m^2\right)+6\,a^4 m^4\left( 7\,E_0^2-2\,m^2\right)
\right.\right.
\\  & \ds  \left.\left.-a\,m\,\mu^2{\left( a\,m +2\, b\, E_0 \right)}
     \left(5\,L\,E_0^2+13\,L\,m^2-6\,m^2\right)\right.\right.
\\  & \ds
     \left.\left.+b^2m^2\mu^2\left(5\,L\,m^2-23\,L\,E_0^2+6\,E_0^2\right)
 \right.\right.
\\   & \ds  \left.\left.
     -6\,b^2{\left( a\,m +b\, E_0
     \right)}^2\left(13\,m^4-28\,m^2E_0^2+5\,E_0^4\right)
 \right.\right.
\\  & \ds
     \left.\left.+6\,b^2 E_0^2\left(22\,a^2 m^2 E_0^2-2\,a^2 m^4
     -5\,b^2 E_0^4\right)
  \right]   \right\}
\\[7pt]
E_5 & \ds   = \frac{{\lambda }^5\,b\,\left( {b\,m + a\, E_0}
\right)\,}{3840\,{\left( a\,m + b\, E_0 \right)}^3\,{\mu
}^4\,m^4}\,
\\& \ds \times
    \left\{
     45\,{\left( a\,m + b\, E_0 \right)}^6
     +105\,b^2\mu^2{\left( a\,m + b\, E_0 \right)}^4 \right.\\[7pt]
     & \ds  \left.
     +L\,\left[ 24\,a^3 m^3 b\,E_0\left(6\,E_0^2-m^2\right)
     +6\,a^4 m^4\left( 7\,E_0^2-2\,m^2\right)
\right.\right.
\\
     & \ds  \left.\left.
     -a\,m\,\mu^2{\left( a\,m +2\, b\, E_0 \right)}
     \left(5\,L\,E_0^2+13\,L\,m^2-6\,m^2\right)\right.\right.
\\
     & \ds
     \left.\left.
     +b^2m^2\mu^2\left(5\,L\,m^2-23\,L\,E_0^2+6\,E_0^2\right)
 \right.\right.
\\
     & \ds  \left.\left.
     -2\,b^2{\left( a\,m +b\, E_0
     \right)}^2\left(19\,m^4-44\,m^2E_0^2-5\,E_0^4\right)
 \right.\right.
\\     & \ds
     \left.\left.+6\,b^2 E_0^2\left(22\,a^2 m^2 E_0^2-2\,a^2 m^4
     -5\,b^2 E_0^4\right)
  \right]   \right\}
\end{array}
\end{equation}
where $\mu^2 = m^2 - E_0^2\,$, $L = l \left( l + 1 \right)$,
$N=n+1/2+\sqrt{{W_0^2-V_0^2+(l+1/2)^2}}$.

It is readily seen that the use of $ \hbar$-expansion technique
leads to the explicit perturbation expansion in powers of the
small parameter $\lambda$.

 Besides, it is known the $s$-wave solution to the
Klein-Gordon
 equation with vector and scalar Hulth\'en-type
potentials, obtained in terms of hypergeometrical functions. From
equation for allowed energy values, listed in \cite{b27}, one can
derive the explicit expression for exact $s$-wave energy
eigenvalues
\begin{equation}\label{4.7}
E=\frac{-a\,
b\,m/\hbar^2+\lambda\,a\;\kappa^2/2+\tilde{N}\,\kappa\,c\sqrt{m^2\,c^2+\lambda
\,m\,b-\lambda^2 \kappa^2\hbar^2/4 }}{\tilde
N^2+a^2/(\hbar\,c)^2}
\end{equation}
where $\tilde
N=n+1/2+\sqrt{b^2/(\hbar\,c)^2-a^2/(\hbar\,c)^2+{1}/{4}}$ and
$\kappa=\sqrt{\tilde N^2 + a^2/(\hbar\,c)^2 - b^2/(\hbar\,c)^2 }$.

With performing the perturbation expansion of this expression in
powers of $\lambda$ we arrive at the system (with $\hbar\,c=1$)
\begin{eqnarray}\label{4.8}
E_0 & = \frac{m}{\hbar^2}\;\frac{ -a\; b+\tilde N\;\kappa}{\tilde
N^2+a^2}\nonumber
\\
E_1 & =\frac{\lambda}{2}\; \frac{\kappa\,(\kappa\;a+\tilde N\;
b)}{\tilde N^2+a^2} \nonumber
\\[7pt]
E_2 & = -\frac{\lambda^2\,\hbar^2}{8\,m}\;\frac{\tilde
N\kappa\,(\kappa^2+b^2)}{\tilde N^2+a^2}
\\E_3 & = -\frac{\lambda\;\hbar^2\;b}{2\,m\,}\;E_2 \nonumber
\\[7pt]E_4 & =
\lambda^2\;\hbar^4\;\frac{\kappa^2+5\,b^2}{16\,m^2\,}\;E_2\nonumber
\\[7pt]E_5 & = -\lambda^3\;\hbar^6\;\frac{b\,(3\,\kappa^2+7\,b^2)}{32\,m^3}\;
E_2\nonumber
\end{eqnarray}
which coincides with (\ref{4.6}) under the condition $l=0$. Note
that for $s$-waves there exists a critical value of the screened
parameter $\lambda$ . From (\ref{4.7}) the obvious condition
\begin{equation}\label{4.9}
m^2c^2+\lambda\;m\,b-\frac{\lambda^2\;\kappa^2\hbar^2}{4}\geq 0
\end{equation}
yields
\begin{equation}\label{4.10}
\lambda_{cr}=\frac{2\,m\,c^2}{\sqrt{\tilde N^2\hbar^2c^2 + a^2} -
b}.
\end{equation}
Then bound states can occur provided the condition $ \lambda <
\lambda_{cr} $.

In order to assess the speed and accuracy of the perturbation
technique for the Hulth\'en potential with various ratios of its
components we consider energy eigenvalues for the pure vector
case, $V(r)= -\frac{a\lambda}{e^{\lambda r}-1}$; the pure scalar
potential, $W(r)=-\frac{b\lambda}{e^{\lambda r}-1}$; and the
equally mixed interaction. Typical results of calculation are
presented in Tables where sums of first
terms from our expansion for the energy eigenvalues for the bound
state problem within the framework of the radial Klein-Gordon
 equation is compared with the results obtained by numerical
integration, $E_{\rm num}$ (in relativistic units $\hbar = c = m
= 1$). As the numerical integration procedure was used, the
improved shooting method with the Noumerov integration scheme,
described in literature \cite{b28}, which ensures the necessary
exactness of calculation for the Sturm-Liouville problems.

The Table 1 demonstrates dependence of accuracy of the
perturbation expansions on the value of the screening parameter
$\lambda$. In Table 1 are shown the results of the calculations
of the sum of first five terms from perturbation expansions for
energy eigenvalues, computed for the first excited $p$-state
($n=1$, $l=1$) with parameters $a=1.0$, $b=1.0$ and the value of
$\lambda$ varying from 0.05 to 0.15. It is seen that the accuracy
of the perturbation description decreases with increasing the
parameter $\lambda$.

The Table 2 illustrates the speed of the perturbation technique on
the energy eigenvalue calculation for the Hulth\'en potential
with various ratios of its component.The sequences of the partial
sums of $k$ corrections to the energy eigenvalues were computed
for the states with $l=1$, $n=1$ and $n=2$, and parameters
$a=1.0$, $b=1.0$, $\lambda=0.05$.

As can be seen from Table 2 the sequences of partial sums of
perturbation corrections to the energy eigenvalues for the
Hulth$\acute{\rm e}$n potential have different behaviour in the
pure vector case and for scalar case and mixed interaction.

Because the parameter $b$, which determines the scalar
interaction, is involved linearly in all perturbation corrections
of odd orders, starting from third (see equations (\ref{4.6})),
these corrections in the pure vector case become equal to zero.
Then we have one subsequence that tends smoothly to the exact
value.

When the pure scalar or mixed interaction is considered, we have
two subsequences bounded below and above the energy eigenvalues.
The average of these subsequences at the point of their maximal
drawing together is proved to result in a quite good
approximation to the exact value.

\section{Summary}
A new useful semiclassical technique for deriving results of the logarithmic
perturbation theory for
 the bound-state problem withing the framework of the
radial Klein-Gordon equation
with the screened Coulomb potential having both time-component of Lorentz-vector
and a Lorentz-scalar term has been
developed. Based upon the $\hbar $-expansions and suitable
quantization conditions, new handy recursion relations have been obtained. Avoiding the
disadvantages of the standard approach these formulae have the
same simple
 form both for ground and exited states and provide,
in principle, the calculation of the perturbation corrections up
to an arbitrary order in the analytic or numerical form. And, at
last, this approach does not imply knowledge of the exact
solution for zero approximation, which is derived automatically.
As an example of application, perturbation expansions for the
energy eigenvalues for the Hult\'en potential containing the
vector part as well as the scalar component have been
investigated with proposed technique. The different behaviour of
the perturbation series in the case of potential in the form of
time-component of a Lorentz-vector and for a Lorentz-scalar
interaction has been found.

\ack This research was supported by a grant N 0100V005251 from the
Ministry of Education and Science of Ukraine which is gratefully
acknowledged.

\eject

\begin{table*}
\begin{center}
\caption{The partial sums of $5$ corrections to the energy
eigenvalues $E_V$ for pure vector Hulth\'en potential, $E_W$ for
pure scalar potential, and $E_{V+W}$ for equally mixed
interaction, and percentage error $\varepsilon$, calculated with
parameters $a=1.0$, $b=1.0$, in units $\hbar=m=c=1$.}
\small{\begin{tabular}{lllllll} \br $\lambda$  &
\multicolumn{1}{c}{$E_V$}& \multicolumn{1}{c}{$\varepsilon_V\%$} &
\multicolumn{1}{c}{$E_W$} & \multicolumn{1}{c}{$\varepsilon_W\%$}
& \multicolumn{1}{c}{$E_{V+W}$} &
\multicolumn{1}{c}{$\varepsilon_{V+W}\%$}
\\ \mr
0.05 & 0.95706870 & 0.00001 & 0.97392119 & 0.00003 & 0.84245453 & 0.00001 \\
0.06 & 0.96113964 & 0.00002 & 0.97739507 & 0.00008 & 0.85034939 & 0.00002 \\
0.07 & 0.96503895 & 0.00005 & 0.98064017 & 0.00020 & 0.85805052 & 0.00004 \\
0.08 & 0.96876527 & 0.00011 & 0.98365749 & 0.00045 & 0.86555969 & 0.00008 \\
0.09 & 0.97231703 & 0.00023 & 0.98644772 & 0.00091 & 0.87287852 & 0.00016 \\
0.1 & 0.97569245 & 0.00044 & 0.98901136 & 0.00171 & 0.88000851 & 0.00030 \\
0.11 & 0.97888956 & 0.00080 & 0.99134866 & 0.00307 & 0.88695104 & 0.00052 \\
0.12 & 0.98190615 & 0.00138 & 0.99345974 & 0.00527 & 0.89370738 & 0.00086 \\
0.13 & 0.98473983 & 0.00229 & 0.99534457 & 0.00873 & 0.90027874 & 0.00137 \\
0.14 & 0.98738798 & 0.00369 & 0.99700303 & 0.01414 & 0.90666626 & 0.00210 \\
0.15 & 0.98984779 & 0.00577 & 0.99843492 & 0.02255 & 0.91287103 & 0.00312 \\
\br
\end{tabular}}
\end{center}
\end{table*}

\begin{table*}
\caption{The sequences of the partial sums of $k$ corrections to
the energy eigenvalues $E_V$ for pure vector Hulth\'en potential,
$E_W$ for pure scalar potential, and $E_{V+W}$ for equally mixed
interaction, calculated with parameters $a=1.0$, $b=1.0$,
$\lambda=0.05$, in units $\hbar=m=c=1$. } \small{
\begin{tabular}{lllllll} \br k&\multicolumn{3}{c}{$n=1$,
$l=1$}&\multicolumn{3}{c}{$n=2$, $l=1$}
\\ \cline{2-7} & \multicolumn{1}{c}{$E_V$}& \multicolumn{1}{c}{$E_W$} &
\multicolumn{1}{c}{$E_{V+W}$} & \multicolumn{1}{c}{$E_V$} & \multicolumn{1}{c}{$E_W$} &
\multicolumn{1}{c}{$E_{V+W}$}
\\ \mr
0 & 0.9341723590 & 0.9530618622 & 0.8000000000 & 0.9638612635 & 0.9726183555 & 0.8823529412 \\
1 & 0.9591723590 & 0.9768884088 & 0.8450000000 & 0.9888612635 & 0.9969338144 & 0.9294117647 \\
2 & 0.9570741392 & 0.9738581555 & 0.8423958333 & 0.9848180151 & 0.9915208448 & 0.9246200980 \\
3 & 0.9570741392 & 0.9739339118 & 0.8424609375 & 0.9848180151 & 0.9916561690 & 0.9247398897 \\
4 & 0.9570686998 & 0.9739202644 & 0.8424540955 & 0.9847983143 & 0.9916167626 & 0.9247202876 \\
5 & 0.9570686998 & 0.9739211933 & 0.8424545273 & 0.9847983143 & 0.9916195489 & 0.9247216080 \\
6 & 0.9570686381 & 0.9739209078 & 0.8424544790 & 0.9847977509 & 0.9916177074 & 0.9247213732 \\
7 & 0.9570686381 & 0.9739209397 & 0.8424544833 & 0.9847977509 & 0.9916179261 & 0.9247213972 \\
8 & 0.9570686368 & 0.9739209276 & 0.8424544828 & 0.9847977150 & 0.9916177253 & 0.9247213921 \\
9 & 0.9570686368 & 0.9739209295 & 0.8424544828 & 0.9847977150 & 0.9916177586 & 0.9247213928 \\
10 & 0.9570686367 & 0.9739209288 & 0.8424544828 & 0.9847977119 & 0.9916177282 & 0.9247213926 \\
 \\
$E_{num}$ & 0.9570686367 & 0.9739209289 & 0.8424544828 & 0.9847977115 & 0.9916177295 & 0.9247213926 \\
\br
\end{tabular}}
\end{table*}

\end{document}